\documentclass[aps,prl,twocolumn,showpacs]{revtex4}
\usepackage{graphicx}\usepackage{subfigure}\usepackage[usenames]{color}
\newcommand{\beq}{\begin{eqnarray}}
\newcommand{\eeq}{\end{eqnarray}}

\begin{document}

\preprint{1}

\title{Light-harvesting in bacteria exploits a critical interplay \\ between transport and trapping dynamics}
\author {Felipe Caycedo-Soler, Ferney J. Rodr\'{i}guez and Luis Quiroga}
\affiliation{Departamento de F\'{i}sica, Universidad de Los Andes,
A.A. 4976 Bogot\'a, D.C.,Colombia}
\author {Neil F. Johnson}
\affiliation{Department of Physics, University of Miami, Coral Gables, Miami, Florida 33126, USA}
\date{\today}
\begin{abstract}
Light-harvesting bacteria {\it Rhodospirillum Photometricum} were
recently found to adopt strikingly  different architectures
depending on illumination conditions. We present analytic and
numerical calculations which explain this observation by
quantifying a dynamical interplay between excitation transfer
kinetics and reaction center cycling. High light-intensity
membranes (HLIM) exploit dissipation as a photo-protective
mechanism, thereby safeguarding a steady supply of chemical
energy, while low light-intensity membranes (LLIM) efficiently
process unused illumination intensity by channelling it to open
reaction centers. More generally, our analysis elucidates and
quantifies the trade-offs in natural network design for solar
energy conversion.
\end{abstract}
\maketitle

Photosynthesis is Nature's solution to the solar energy conversion
problem \cite{fcondon,sension,Geyer,review,fleming_r}.
Understanding what architectural designs it adopts, and why, are
important questions which could guide
designs of future energy conversion devices. Many studies
have clarified the exciton capture-transfer dynamics of reaction center (RC) pigment-protein
complexes\cite{review,fleming_r,Silbey,lloyd1,ritz,sumi}, their
arrangement along the membranes that support them
\cite{bahatyrova,sturg1,sturgdiff,gonzalves!,scheuring2,scheuring}, and even
RC quantum effects\cite{flemming,Engel}. Recent experimental
investigations \cite{sturg1} resolved the locations of Light
 Harvesting (LH) complexes within the 2D membrane architecture of complete chromatophore
 vesicles, to reveal an unexpected change
in the ratio of complexes for bacteria grown under high
(Fig. 1(a)) versus low illumination intensities (Fig. 1(b)).\\

This Letter presents analytic and numerical results which explain
this experimental observation \cite{sturg1} by quantifying a
trade-off which arises between two fundamental membrane
requirements: (1) the need to convert large numbers of excitations
into energetically useful charge separations within the RC, and
hence promote metabolic activity, and (2) the need to avoid an
oversupply of excitations and hence excessive bursts of energy,
which could damage the photosynthetic machinery
\cite{sturg1,sturgdiff}. Within our theory, the microscopic origin
of this trade-off is the interplay between excitation transfer
kinetics across the membrane architecture and reaction center
cycling dynamics. This generates a critical behavior of the
membrane's efficiency when probed under different light
intensities.  Low light-intensity membranes (LLIMs) efficiently
channel excess illumination intensity to open reaction centers,
and hence are dominated by (1), while   high light-intensity
membranes (HLIMs) better exploit  excitation loss through
dissipation as a photo-protective mechanism in order to provide
constant chemical energy, and hence are dominated by (2). Our
analytic model predicts a critical light intensity during growth,
below which the synthesis of  LH2 complexes should be dramatically
enhanced.

The photosynthesis process in purple bacteria \cite{review,
fleming_r}  (see Fig.1(c)) involves photons from sunlight being
absorbed with a rate
$\gamma_{A}=I(\gamma_{1}N_{1}+\gamma_{2}N_{2})$, where $N_{1(2)}$
is the number of LH1 (LH2) complexes, $\gamma_{1(2)}$\cite{Geyer}
are the respective absorption rates per complex, and $I$ is the
light intensity. Each LH1 contains one RC, i.e. the number of RCs
is equal to $N_1$. Excitation transfer between light harvesting
complexes occurs  at the following mean times (picoseconds)
\cite{ritz,flemming,sumi}: $t_{12}=15$ for LH1-LH2; $t_{21}=3.3$
for LH2-LH1; $t_{11}=20$ for LH1-LH1; $t_{22}=10$ for LH2-LH2;
before reaching an RC at time $t_{1,RC}=25$ for LH1-RC. Once in an
open RC, the special pair may become ionized $(P^{+})$ on the
timescale $t_{+}=3$, eventually producing quinol (Q$_B$H$_2$) by
reducing quinone (Q$ _B\rightarrow$Q$ _B^-\rightarrow$Q$ _BH$)
twice \cite{fcondon}. Otherwise back transfer occurs from RC to
LH1 with $t_{RC,1}=8$. Before a new Q$_B$ becomes available, the
RC remains closed  for a cycling time $\tau$ during which an
energetically useful charge separation is generated
\cite{Geyer,sturgdiff,RC1,RC2}. Dissipation through fluorescence
or internal conversion happens at a constant rate
$\gamma_D$\cite{Geyer}.
\begin{figure}
\begin{center}
\includegraphics[width=8.0 cm] {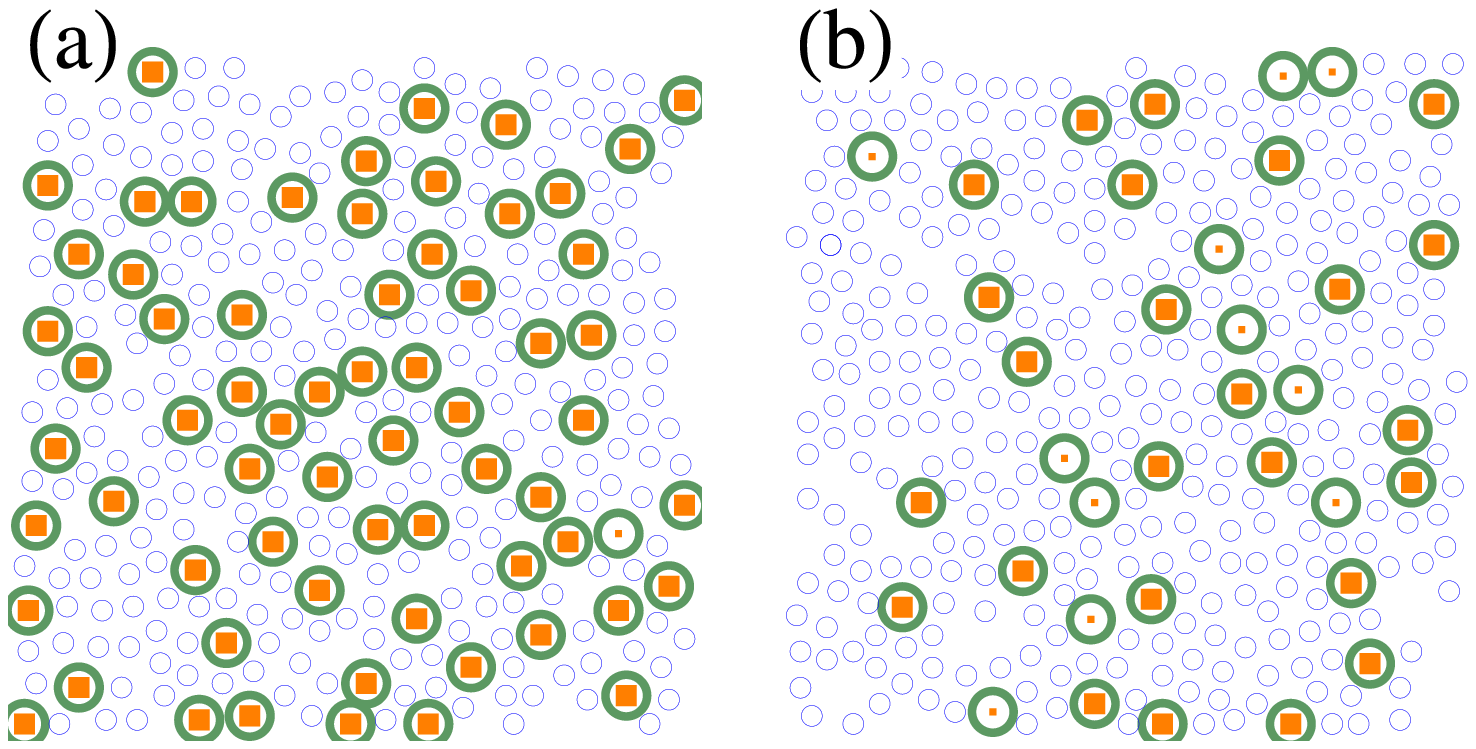}
\includegraphics[width=8.0  cm]{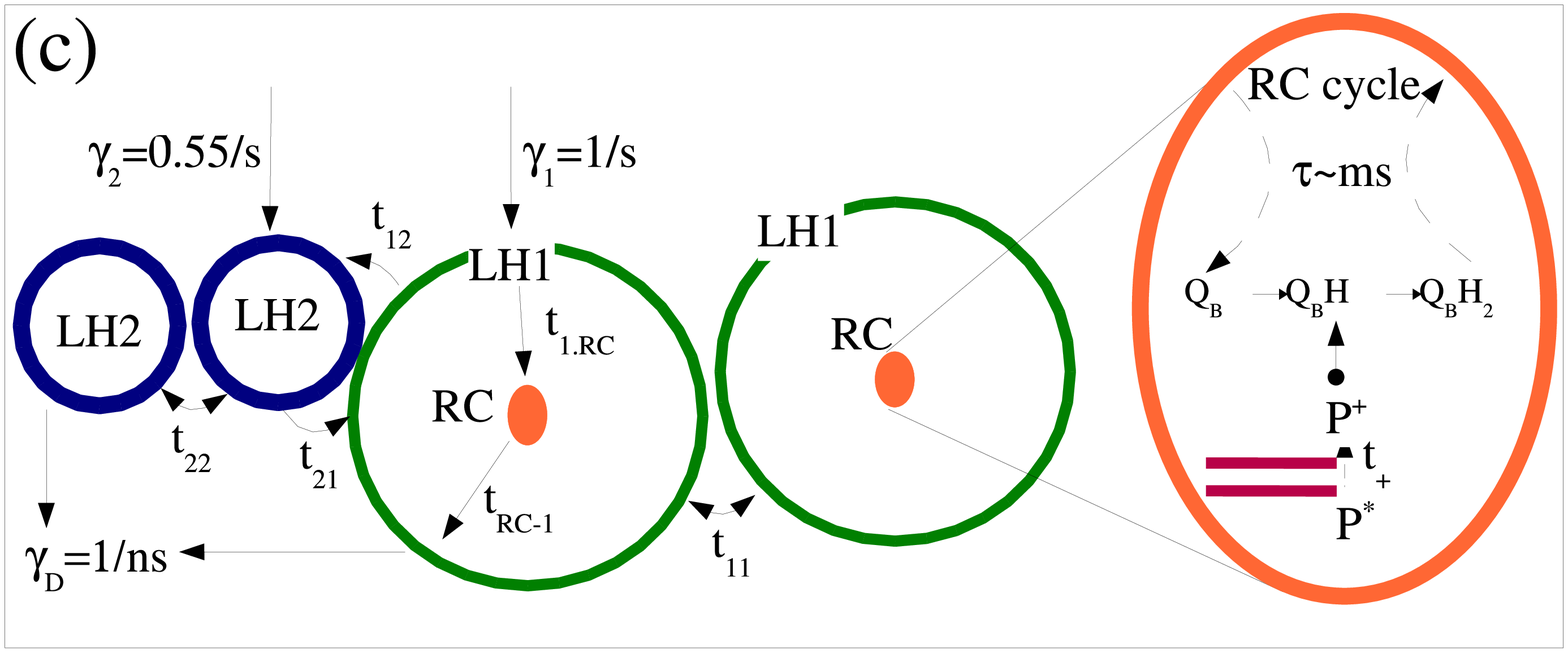}
\vspace{-3 cm}
\end{center}
 \caption{(Color online) Top panels: Empirical
  architectures  (i.e. digitized positions of LH complexes)
 from Ref. \cite{sturg1}. (a) High light-intensity membrane (HLIM), $I_0=100$ W/m$^2$.
 (b) Low light-intensity membrane (LLIM), $I_0=10$ W/m$^2$. Small orange dots: open
 reaction centers (RC) at snapshot during simulation. Large orange boxes: closed RCs.
 Large green circles: LH1s. Small blue circles: LH2s. (c) Summary of dynamical
 processes of excitation transfer between LH1-LH1,
LH2-LH2, LH1-LH2, LH1-RC. Dissipation ($\gamma_D$), absorption
$(\gamma_{1(2)})$ and RC cycling time $\tau$ (enlarged orange
oval) are also shown.} \label{fig0}
\end{figure}
The photo-excitation kinetics can be described by a collective
population state vector $\rho(t)$ which follows a master equation
$\partial_t\rho_i(t)=\sum_jG_{ij}\rho_j(t)$, where the element
$G_{ij}$ of the rate matrix establishes the probability per unit
time of a transition (due to absorption, dissipation, RC
ionization, transfer to neighbors) between collective states $i$
and $j$. Due to small absorption rates, the probability that two
excitations occupy either a single harvesting structure or RC is
negligible.  Light harvesting complexes only have two possible
states: no exciton present (unexcited) or one exciton state
(excited). However the RC has four possible states: (un)excited
while being open or closed. Hence the state-space has size
$2^{N_1}2^{N_2}4^{N_1}=2^{3N_1+N_2}$.

\begin{figure}
\begin{center}
\includegraphics[width=4.cm]{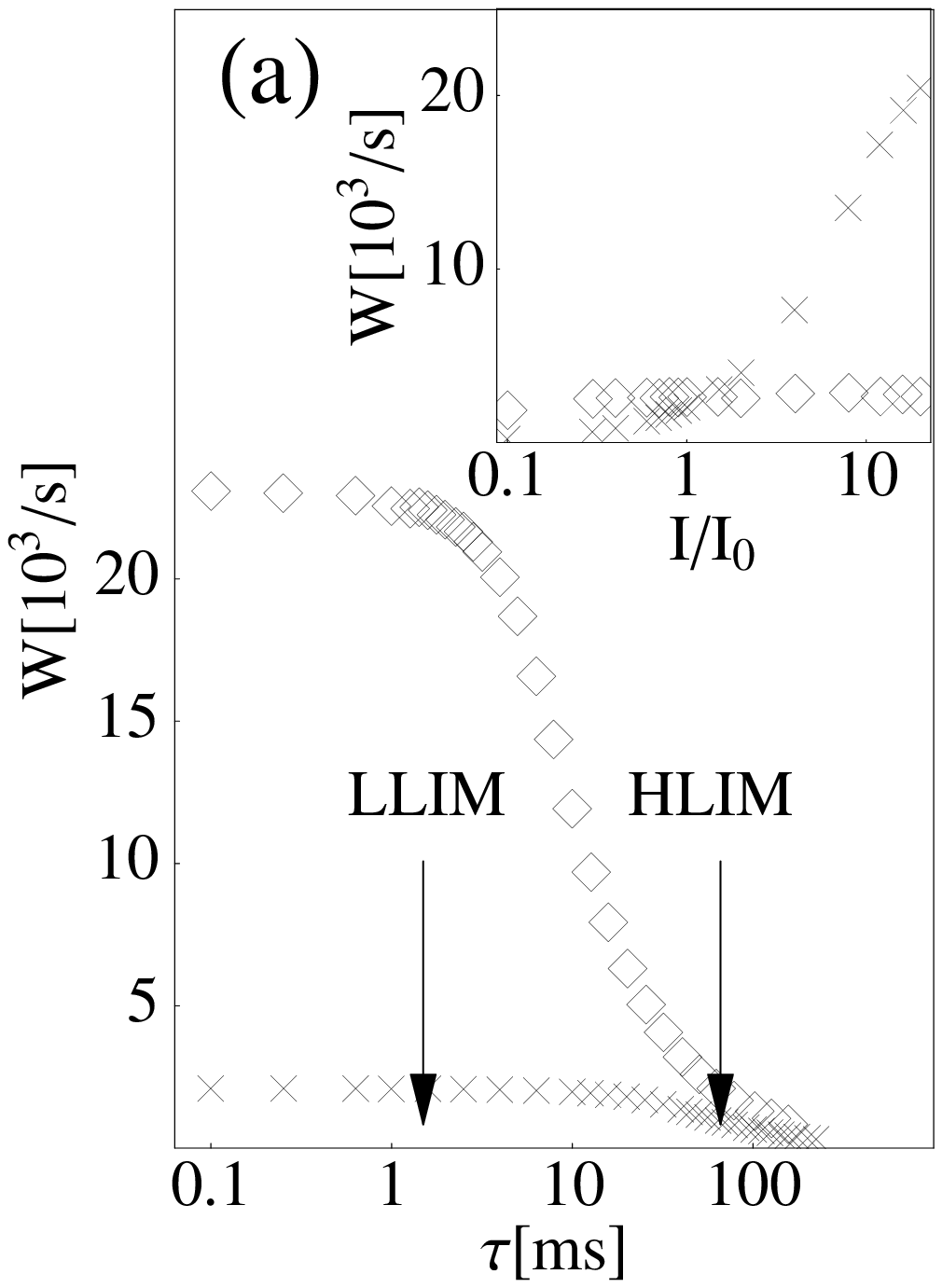}
\includegraphics[width=4 cm]{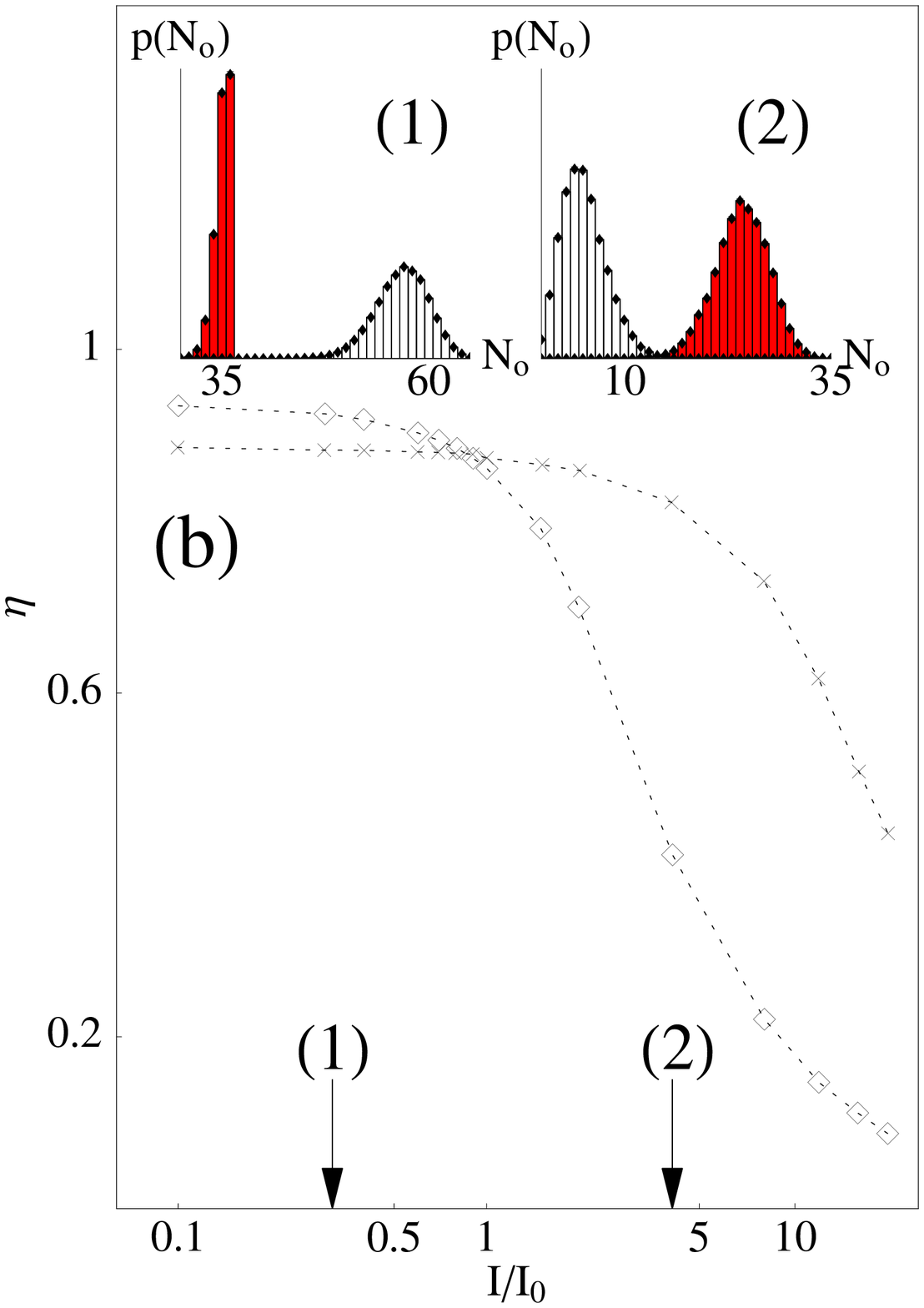}
\vspace{-0.5 cm}
\end{center}
\caption{ (Color online) (a) $W(\tau)$ for LLIM (crosses) and HLIM (diamonds).
The inset shows the quinol rate with respective RC cycling
times for which LLIM has the same quinol output as HLIM, shown
with arrows in the main plot. (b) Efficiency $\eta$ as a function
of normalized intensity $I/I_0$ for $\tau=3$ ms in HLIM (diamonds)
and LLIM (crosses) adapted membranes. Dotted lines are aids to the
eye. In insets, distribution of open RCs, $p(N_{o}),$ in HLIM
(white bars) and LLIM (red bars) adapted membranes,
are shown for the values highlighted by arrows in main plots. Error bars
smaller than the symbols are not shown.}  \label{fig1}
\end{figure}

Vesicles imaged with experimental Atomic Force Microscopy (AFM) \cite{sturg1}
 show that typically $2N_1+N_2\approx 300-600$. Given the large state-space, we study a discrete-time random walk
simulation for excitations which includes dynamical coupling to
open/closed RC states, in contrast to Refs.\cite{review,ritz}. We
use the empirical architectures \cite{sturg1} to establish the
likely neighbors (e.g. within 30$\AA$) between which excitations
can hop, and implement the process of absorption and excitation
 transfer, dissipation, or RC capture (if within a RC) using Monte Carlo.
 All processes obey exponential distributions with mean values presented in
 Fig.1(c). When two excitations reach a single RC, it closes for the cycling time $\tau$
  \cite{Geyer,sturgdiff,RC1,RC2}.
We checked the accuracy of our stochastic numerical simulation by
comparing its population-level predictions to those of a master
equation for small chromatophores \cite{fcs}. The HLIM and LLIM
architectures (Fig. 1, top panel) differ in the relative number of
complexes and have stoichiometry $s$=$N_2/N_1$. A typical snapshot
of open and closed RCs from our stochastic simulations (see Fig. 1
(a) and (b)) demonstrates that HLIM has fewer open RCs than LLIM
in the experimentally relevant regime of millisecond RC cycling
times \cite{RC1,RC2}. Our interest is the actual quinol output of
the membrane, hence we calculate stationary-state observables when
the numerical simulations converge to  a constant quinol rate. The
quinol production rate $W=\frac{1}{2}\frac{d n_{RC}}{dt}$ is half
the rate at which $n_{RC}$ excitations produce  ionization $P^+$.
Assuming similar metabolic requirements under different
illumination growth, the times suggested by arrows in Fig.2(a)
imply that LLIM has a shorter RC-cycling time than HLIM. This is
consistent with greater quinone availability in LH1 clusters, as
appropriate in LLIMs \cite{sturgdiff}.

We find qualitatively different behaviors in $W$ as a
function of normalized intensity $I/I_0$  (see Fig. 2(a)
inset).  In HLIM, greater intensity does not change the quinol
rate, while in LLIM higher illumination  increases $W$. Therefore
for higher light intensity,  LLIM will be better than HLIM at
processing potentially dangerous occurrences of excess  excitations. Due to
fewer open
 RCs, HLIM will process only the necessary number of excitations for metabolism.
The efficiency $\eta=n_{RC}/n_{A}$ is  related to the quinol
rate in the stationary limit
 through $\eta=2W/\gamma_A$, and quantifies the performance of
  a membrane in initiating RC ionizations from the $n_A$ total absorbed
  excitations. Figure \ref{fig1}(b) shows that increased
  light intensity lowers $\eta$ in both membranes due to a reduced
  number of open RCs ($N_0$),
   as shown by the distributions $p(N_o)$ of open RCs  at the top of Fig.2(b).
Consequently  LLIMs have better efficiency than HLIMs since they
have more open RCs in the high light intensity range, even though they have
fewer RCs.

\begin{figure}
\begin{centering}
\includegraphics[width=5.5 cm,height=3.2 cm]{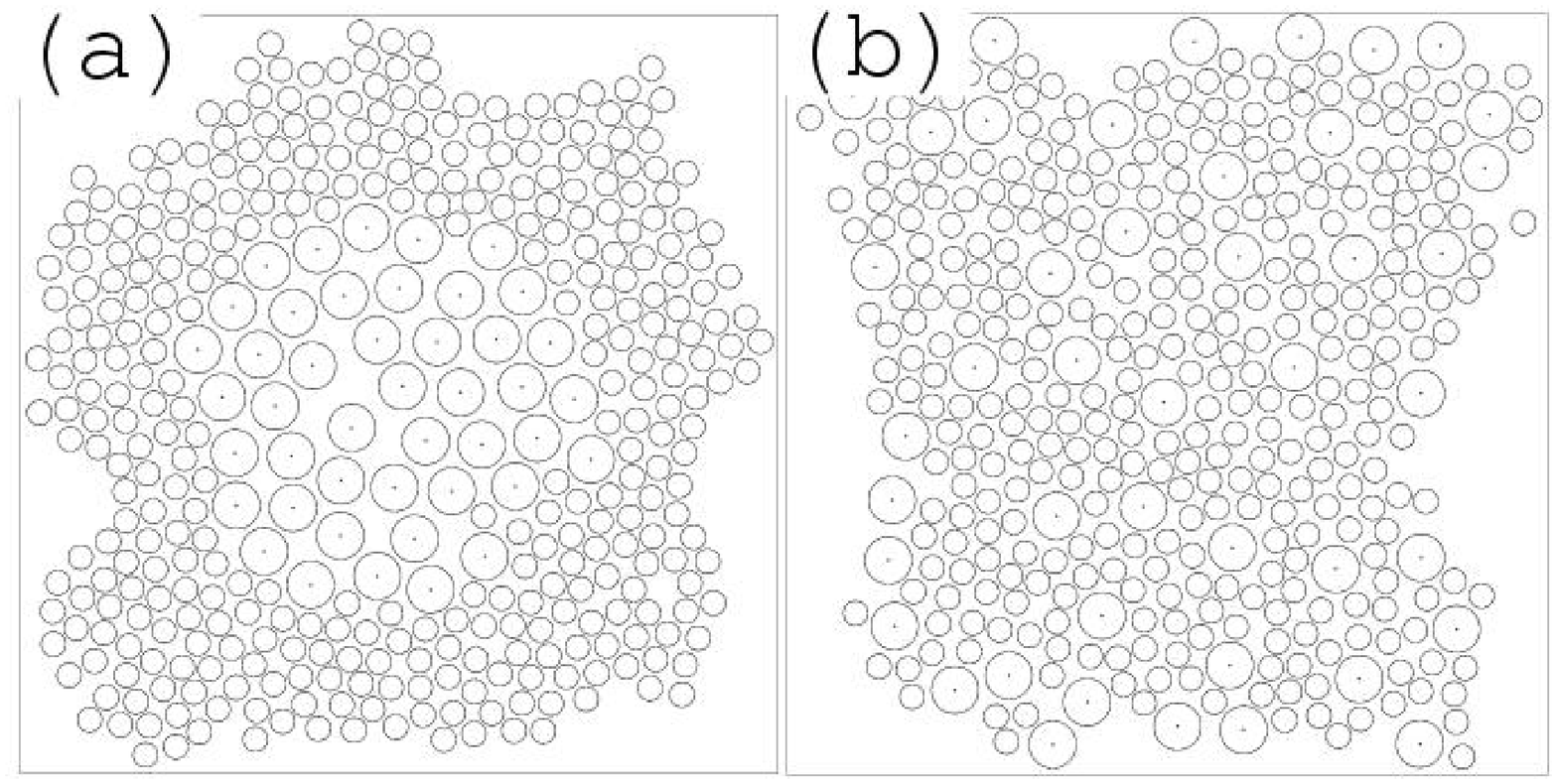}
\hspace{-0.3 cm}
\includegraphics[width=3. cm, height=3.2 cm]{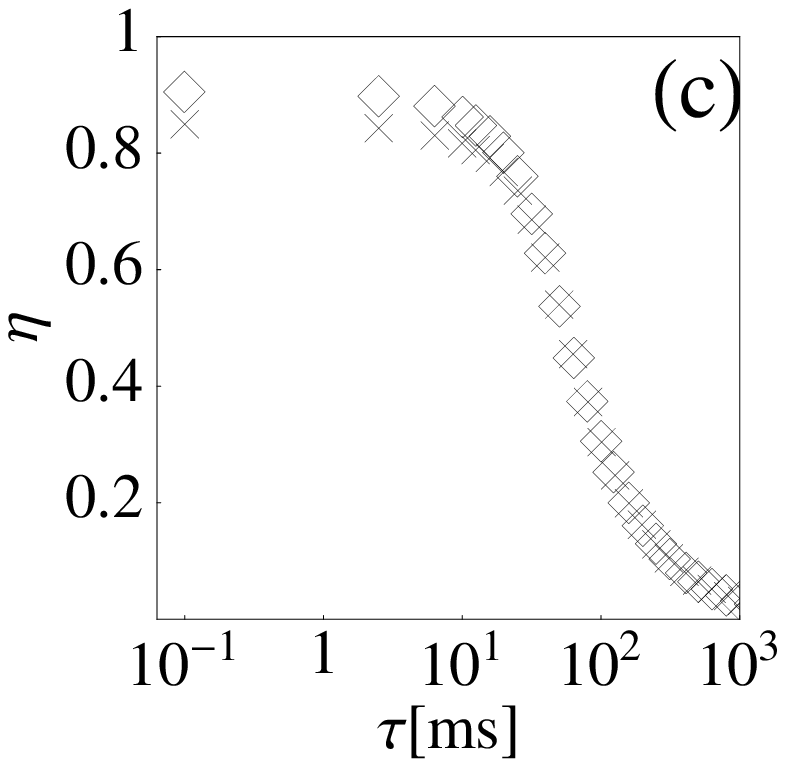}\\

\caption{(a) Ordered and (b) random membranes with $s$=8.09.
(c) presents $\eta(\tau)$ for ordered (crosses) and random
(diamonds) membranes.}\label{fig2}
\end{centering}
\end{figure}

An intriguing question arises as to whether clustering of LH1s might help reduce the
effective path that an excitation needs to take to a closed RC \cite{sturg1}.
To explore this, we consider two extreme
chromatophore vesicles (Fig. \ref{fig2}(a) and (b)), both of which
are compatible with relevant LLIM stoichiometries
\cite{scheuring2}. Their efficiencies are shown in Fig.
\ref{fig2}(c).   In the experimentally relevant regime of
millisecond RC cycling-time, open RCs are sparse -- they are not clustered, and the
architecture has no
significant influence on $\eta$.  Less
clustering does induce a slightly higher efficiency since the RC
borders become easier to reach \cite{ritz,fcs}. However apart from the benefit
of quinone exclusion, clustering seems not to
appreciably diminish the path length of excitations reaching a
closed RC. On the other hand, clustering of LH2s
might exclude active quinones from LH1-RC domains, thereby increasing
their availability in core clusters and decreasing the RC-cycling
time. Efficiency of the membranes will also depend on $\tau$,
since this dictates how fast an RC opens due to a new quinone
being available.  If $\tau$ is taken as a parameter, very small
differences appear between membranes having equal
stoichiometries but different network architectures. In Fig. 4(a),
we compare the efficiencies
 of the representative architectures presented in Fig. 1, calculated
 using the numerical stochastic simulations, as a function of
RC-cycling time. In their respective illumination regimes, LLIM
dissipates less excitations for $\tau$ values in the biologically relevant
millisecond range\cite{RC1,RC2}. This implies that even if their
RC-cycling times are equal (i.e. no enhanced quinone diffusion in
clustered LLIM), LLIM is efficient ($\eta\approx 85\%$) while HLIM
provides a steady quinol supply by exploiting dissipation
($\eta\approx 20-40\%$). These findings hold irrespective of any
other diffusion enhancements.

\begin{figure}
\begin{centering}
\hspace{-1.0 cm}
\includegraphics[width=5.5 cm,height=4.5 cm]{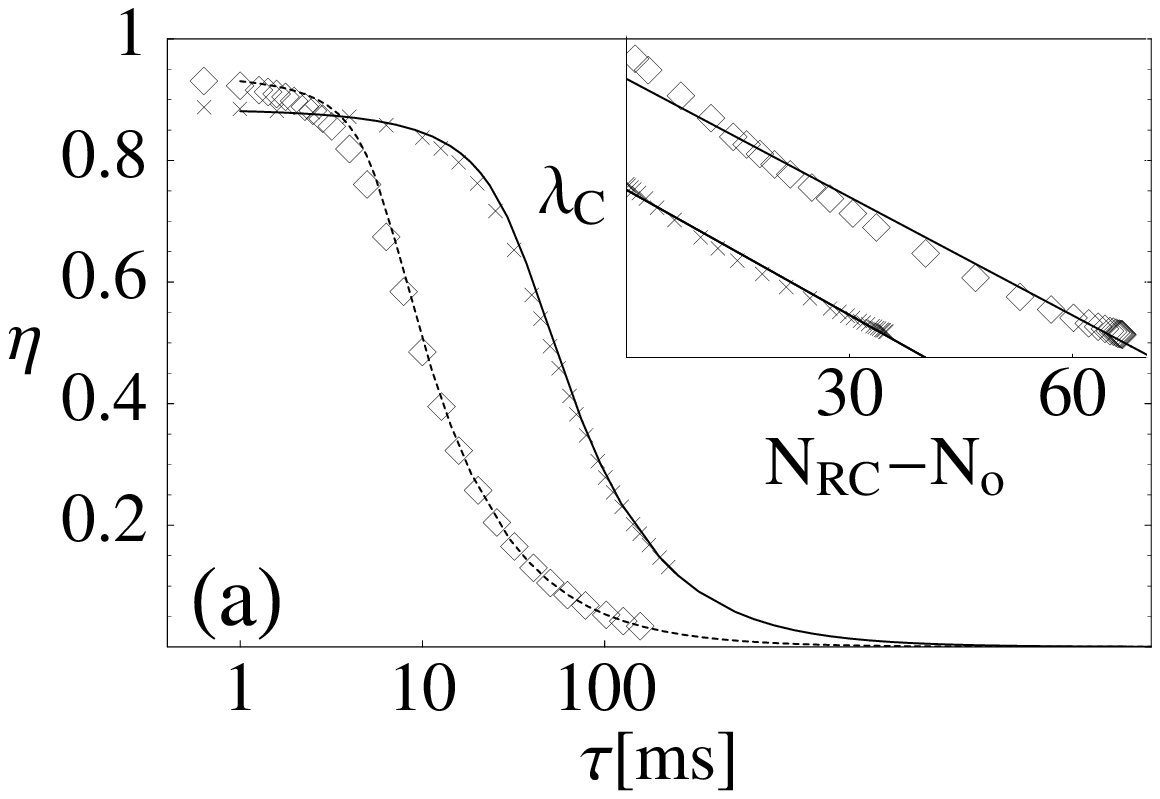}
\hspace{-.3cm}
\includegraphics[width=3.8 cm, height=4.5 cm ]{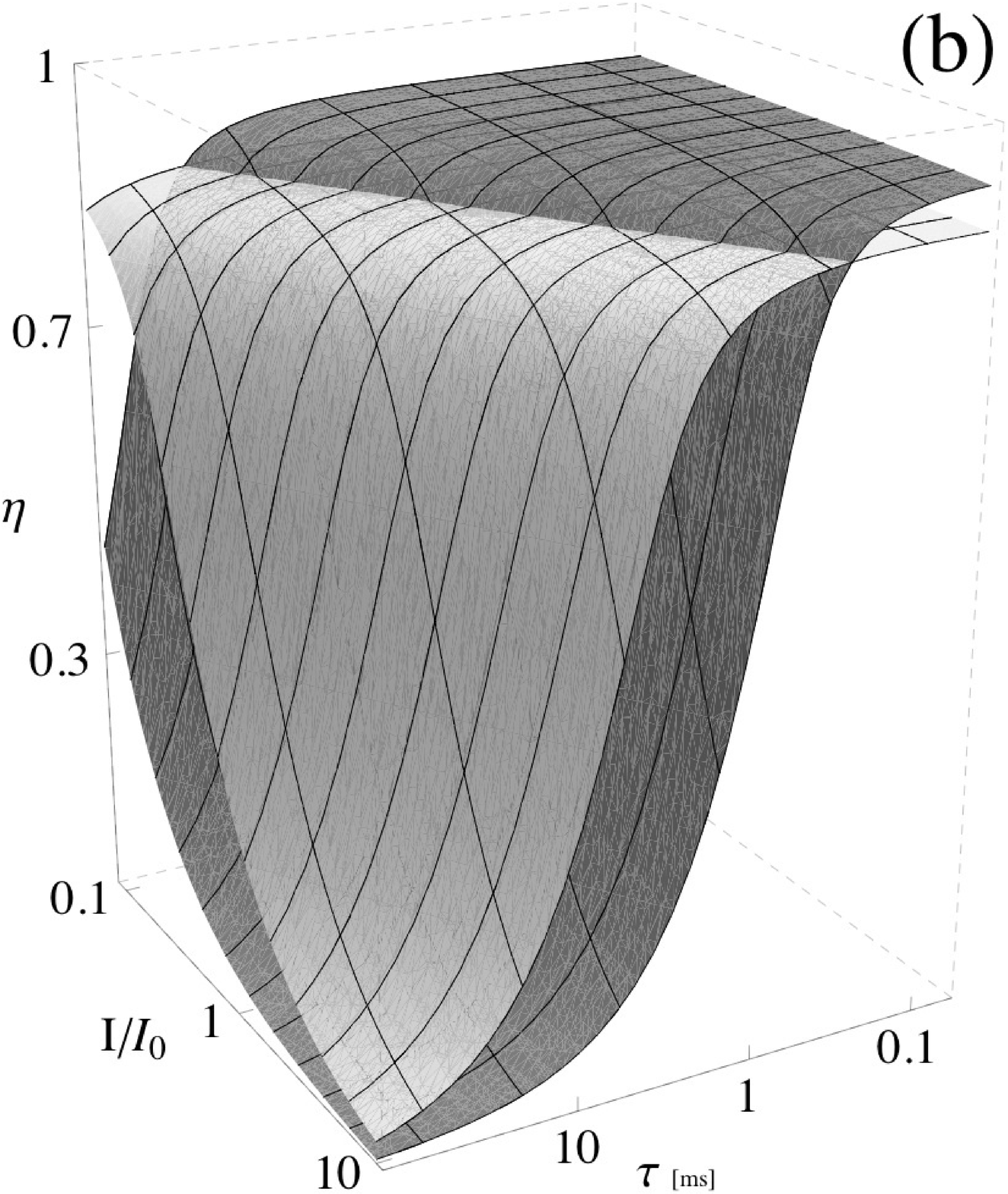}
\hspace{-1.2 cm}
\includegraphics[width=3.5 cm]{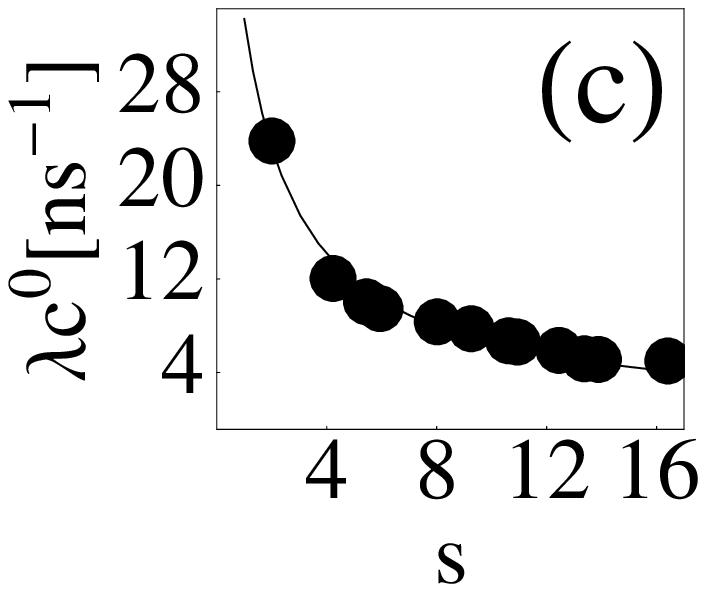}
\includegraphics[width=5.0 cm]{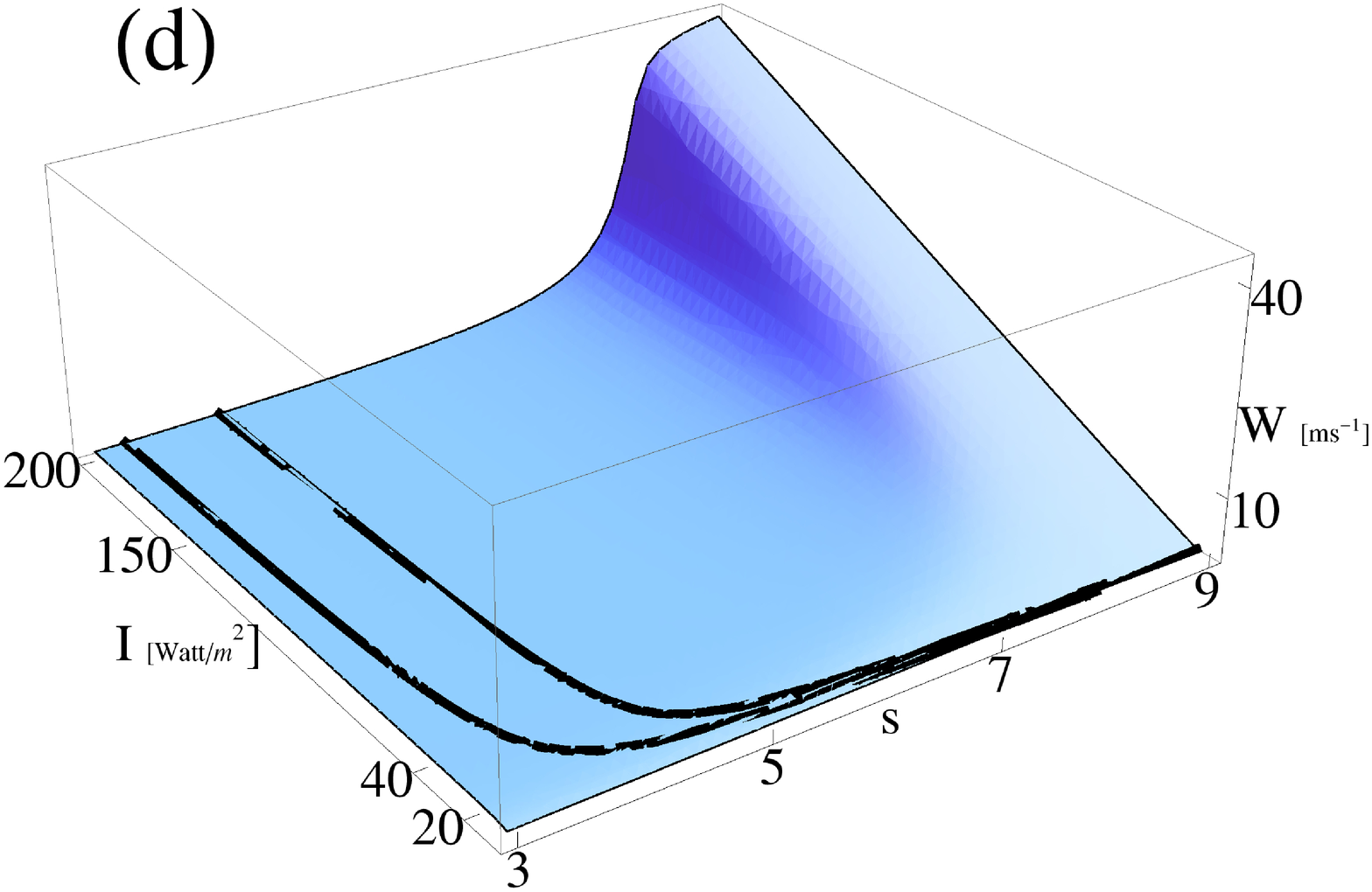}
\caption{(Color online) (a) Analytical model (continuous) and stochastic
simulation (HLIM: diamonds, LLIM: crosses). Inset:
$\lambda_C(N_{o})$  from simulations within the linear portion of the main plot. Parameter values for
LLIM  and HLIM are $\lambda_C^0$=\{0.00771,0.0163\}ps$^{-1}$ and
$N_{1}=\{40.01,70.66\}$ respectively, consistent with empirical
values.  (b)  $\eta$ as function of $\tau$ and $I/I_0$, obtained
from complete analytical solution for LLIM (white) and HLIM
(grey). (c) Numerical calculation of $\lambda_C^0(s)$ (dots) vs.
our analytic form from text (continuous). (d) $W(s,I)$ as function
of stoichiometry $s$ and illumination intensity, with
quinol rate contours of 1900 s$^{-1}$ and 2100
s$^{-1}$.}\label{fig3}
\end{centering}
\end{figure}

Guided by these numerical findings, we now develop an analytic
model to capture the underlying physics. Assume that $N_E$
excitations are created in the membrane at an absorption rate
$\gamma_{A}$. Excitations either leave the membrane through
dissipation at a rate $\gamma_D$,  or at an RC at rate $\lambda_C$
where $\lambda_C$  depends on $N_{o}$. The number of RCs closing
per unit of time is $\lambda_C N_E/2$, while the number opening
per unit of time is $\frac{1}{\tau}(N_1-N_{o})$. Hence the number
of absorbed photons is connected to the number of available (open)
RCs by the following pair of  coupled differential equations:
\begin{eqnarray}
\frac{dN_E}{dt}&=&-(\lambda_C(N_{o})+\gamma_D) N_E+\gamma_{A}  \label{ne}\\
\frac{dN_{o}}{dt}&=&\frac{1}{\tau}(N_1-N_{o})-\frac{1}{2}\lambda_C(N_{o})
N_E \label{noff}
\end{eqnarray}
When $\lambda_C=0$, all RCs are closed. The maximum value
($\lambda_C^0$) occurs when all RCs are open.  When the membrane
is excited, the transfer-ionization rate per open RC is constant
due to the fast excitation hopping relative to the cycling time,
i.e. $\lambda_C/N_o=\lambda_C^0/N_1$ which is also supported by
numerical simulations (see Fig.4(a) inset). Given that
$\eta=\lambda_C^0 N_o N_E/(N_1\gamma_A)$, the steady state
solution to Eqs. (\ref{ne}) and (\ref{noff}) is \cite{fcs}:
\begin{eqnarray}
\eta&=&\frac{1}{{2\gamma_A\lambda_C^0\tau}}
\left\{2N_{1}(\lambda_C^0+\gamma_D)+\gamma_A\lambda_C^0\tau-
    \left[4N_{1}^2(\lambda_C^0+\gamma_D)^2\nonumber\right.\right.\\
& &\left.\left.
+4N_{1}\gamma_A\lambda_C^0(\gamma_D-\lambda_C^0)\tau+(\gamma_A\lambda_C^0\tau)^2\right]^{1/2}\right\}
\end{eqnarray}
In the limit of fast RC cycling-time ($\tau$$\rightarrow$0),
$\eta$ has the simple form $\eta=(1+\gamma_D/\lambda_C^0)^{-1}$.
If all transfer paths are summarized by $\lambda_C^0$, this
solution illustrates that $\eta \geq 0.9$ \cite{ritz}, if the
transfer-P reduction time is less than one tenth of the
dissipation time in the absence of RC cycling. For  finite $\tau$
this analytic solution is in very good quantitative agreement with
the numerical stochastic simulation, supporting our previously
discussed interpretation (see Fig. \ref{fig3}(a)). Figure
\ref{fig3}(b) shows the complete analytical solution of Eqs.
(\ref{ne}) and (\ref{noff}), in order to confirm the entire range
of light intensities and RC closing times for which LLIM has a
higher efficiency than HLIM. The assumed linear fit for
$\lambda_C$ smears out an apparent power-law behavior. We have yet
to find analytical solutions for $\eta$ in cases where
$\lambda_C(N_{o})$ has a power-law dependence.

Our analytical model can be used for easy comparison of
the metabolic  outputs from experimentally distinct AFM-imaged
membranes, in order to provide additional insight concerning the adopted and
expected stoichiometries in {\it Rsp.
Photometricum}\cite{sturg1}.  The bacteria are studied under
different illumination conditions, assuming  that comparable
metabolic needs (i.e. quinol supply) are accomplished in vesicles
of area $A_0$. Our present aim is to find an expression for the quinol
production rate $W$ in terms of the environmental growth
conditions and the responsiveness of purple bacteria through
stoichiometry adaptation.  In the stationary state,
$W=\lambda_C N_E/2$ depends on the number of excitations within
the membrane and on the details of transfer  through the rate
$\lambda_C$.
The LH1 and LH2 complexes of area $A_1$ and  $A_2$,
respectively, fill a fraction $p$ of the total vesicle area
$p=(A_1N_1+A_2N_2)/A_0$. This surface occupancy has been shown
\cite{sturgdiff} to vary among adaptations, since LLIMs have a
greater occupancy ($p\approx 0.85$) than HLIMs ($p\approx 0.75$)
due to para-crystalline domains. The mean number of open RCs in the
stationary state is
$N_{o}=N_1-\frac{\lambda_C\gamma_A}{2(\gamma_D+\lambda_C)}\tau=N_1-W\tau$.
The linear  $\lambda_C(N_o)$ assumption gives $
\lambda_C(s,W)=\lambda_C^0(s)\left(1-\frac{W\tau(s)(A_2s+A_1)}{A_0\
p(s)}\right)$. The RC cycling time $\tau(s)$ is
expected to vary somewhat with adaptations due to quinone
diffusion and different metabolic demands, and is described with a linear
interpolation using the values highlighted by arrows in Fig.
\ref{fig1}(a). Likewise the rate $\lambda_C^0(s)$ must be zero
when no RCs are present ($s$$\rightarrow$$\infty$), and takes a
given value $\langle t_0\rangle^{-1}$ when the membrane comprises
only LH1s ($s$=0). Its dependence on $s$ is satisfied by the form
$\lambda_C^0(s)=(s/a+\langle t_0\rangle)^{-1}$, with adjustable
parameter $a$, for several computer generated membranes (see Fig.
\ref{fig3}(c)). Solving Eqs.(\ref{ne}) and (\ref{noff}) in the
steady state, we obtain
$W=\frac{\lambda_C(s,W)\gamma_A(s,I)}{2(\lambda_C(s,W)+\gamma_D
)}$ which can be solved to yield:
\beq & &2W(s,I)=\frac{\gamma_A(s,I)}{2}+ \frac{1}{B(s)}
\left(1+\frac{\gamma_{D}}{\lambda_c^0}\right)\\
&+&
\sqrt{\left(\frac{\gamma_A(s,I)}{2}+\frac{1}{B(s)}\left(1+\frac{\gamma_{D}}{\lambda_c^0}\right)\right)^2+\frac{\gamma_A(s,I)}{2B(s)}}\nonumber
\eeq where $B(s)=\frac{\tau(s) (A_1 +s A_2)}{p(s)A_0}$. As can be
seen in Fig.4(d) in the high stoichiometry/high intensity
regime, too many excitations would dangerously increase the
cytoplasmic pH \cite{review,fcondon,Geyer}. Longer cycling times
at higher light intensities are therefore helpful in order to keep
power output bounded. The contours in Fig. \ref{fig3}(d) of
constant quinol production rate $W$, show that only in a very
small intensity range will bacteria adopt stoichiometries
which are different from those experimentally observed in {\it
Rsp. Photometricum} (s$\approx $4 and s$\approx$8) \cite{sturg1}.
The empirical finding in Ref. \cite{sturg1} that membranes with $s$=6 or
$s$=2 are not observed, is consistent with our theory. More
generally, our results predict a great sensitivity of
stoichiometry ratios for 30-40 W/m$^2$, below which membranes
rapidly build up the number of antenna LH2 complexes. This
prediction awaits future experimental verification.

In summary, our analytic and numerical calculations elucidate and quantify the
interplay which arises between local (RC cycling) and
extended dynamics (excitation transfer) in a chromatophore
light-harvesting vesicle. In addition to explaining structural differences
during growth, this new quantitative understanding
may help accelerate development of novel solar micropanels
mimicking natural designs.

F.C.-S. acknowledges financial support from Research Project
Funds of Facultad de Ciencias, Universidad de Los Andes, and
Fundaci\'on Mazda. We are grateful to S. Scheuring and J. Sturgis
for detailed discussions. N. F. J.
acknowledges F. Fassioli and A. Olaya-Castro for initial discussions.



\vspace{-.5 cm}
\begin{thebibliography}{2}
\bibitem{fcondon} H. van Amerongen, L. Valkunas and R. van Grondelle, {\it Photosynthetic Excitons} (World Scientific Publishing Co., Singapore, 2000).
\bibitem{sension} R. J. Sension, Nature {\bf 446}, 740 (2007).
\bibitem{Geyer} T. Geyer and V. Helms, Biophys. Journ., {\bf 91}, 927 (2006).
\bibitem{review} X. Hu, et al.,  Quart. Rev. of Biophys. {\bf 35}, 1 (2002).
\bibitem{fleming_r} Y.-C. Cheng and G.R. Fleming, Annu. Rev. Phys. Chem. {\bf 60}, 241 (2009).
\bibitem{Silbey} S. Jang, M. D. Newton and R.J. Silbey,  Phys. Rev. Lett. {\bf 92}, 218301
(2004).
\bibitem{lloyd1} M. Mohseni et al., J. Chem. Phys. {\bf 129}, 174106 (2008).
\bibitem{ritz} T. Ritz, S. Park, and K. Schulten,  J. Phys. Chem. B {\bf 105}, 8259
(2001).
\bibitem{sumi}  H. Sumi, J. Phys. Chem. B {\bf 103}, 252 (1999). 
\bibitem{sturg1} S. Scheuring and J. Sturgis, Science {\bf 309}, 484 (2005). 
\bibitem{sturgdiff} S. Scheuring and J. Sturgis,  Biophys. Jour. {\bf 91}, 3707 (2006); F. Fassioli et al., Biophys. Jour. {\bf 97}, 2464 (2009).
\bibitem{bahatyrova} S. Bahatyrova, et al.,  Nature {\bf 430}, 1058 (2004). 
\bibitem{scheuring} S. Scheuring, J. Busselez, and D. Levi, J. Biol. Chem. {\bf 280}, 1426 (2005). 
\bibitem{scheuring2} S. Scheuring, et al.,  J. Mol. Biol. {\bf 358}, 83 (2006).
\bibitem{gonzalves!} R. P. Goncalves et al.,  J. Struct. Biol. {\bf 152}, 221 (2005). 
\bibitem{Engel} T. Brixner et al., Nature {\bf 434}, 625 (2005); G.S. Engel et al., Nature {\bf 446}, 782 (2007).
\bibitem{flemming} R. Agarwal, et al., J. Phys. Chem. A {\bf 106}, 7573 (2002). %
\bibitem{RC1} O. Savoth and P. Maroti,  Biophys. Journ. {\bf 73}, 972 (1997). 
\bibitem{RC2} F. Milano et al., Eur. Journ. Biochem. {\bf 270}, 4595 (2003).
\bibitem{fcs} F. Caycedo-Soler, PhD Thesis, Universidad de Los Andes (2010): http://fimaco.uniandes.edu.co/investigacion.html

\end{thebibliography}
\end{document}